\documentclass[elsart12,epsf,eqsecnum,epsfig,graphics,cite]{revtex4}
\def\beq{\begin{equation}}

\def\eeq{\end{equation}}

\begin{document}


\voffset1.5cm

\title{Angular Correlations in Gluon Production at High Energy.}
\author{ Alex Kovner$^{1}$ and Michael Lublinsky$^{2}$}

\affiliation{
$^1$ Physics Department, University of Connecticut, 2152 Hillside
Road, Storrs, CT 06269-3046, USA\\
$^2$Physics Department, Ben Gurion University of the Negev, Beer Sheva 84105, Israel\\}
\date{\today}

\begin{abstract}
We present a general, model independent argument demonstrating that gluons produced in high energy hadronic collision are necessarily correlated in rapidity and also in the emission angle. The strength of the correlation depends on the process and on the structure/model of the colliding particles. In particular we argue that it is strongly affected (and underestimated) by factorized approximations frequently used to quantify the effect.
\end{abstract}
\maketitle
\section{Introduction}
The CMS observation of angular and long range rapidity correlations in the hadron spectrum, the so called "ridge" in proton-proton collisions\cite{cms},  has triggered a lot of discussions in recent literature \cite{recent},\cite{ddgjlv}. A similar if more pronounced correlated structure was observed in gold-gold collisions at RHIC \cite{rhicridge}. There is a variety of candidate explanations for the RHIC observation \cite{rhicexplane}, \cite{glv} most of them utilizing strong radial flow as a collimating mechanism. Although flow measurements have not been reported for the LHC data, it is difficult to imagine that flow will have a significant effect in p-p collisions. Thus the viable explanation should probably not appeal to any collective behavior of produced particles.

The purpose of this note is to point out that at high energy, rapidity and angular correlations between produced particles are to be expected on very general grounds. The framework of our discussion here is similar to that of \cite{ddgjlv}, but the argumentation will be quite general without referring to specific models of high energy evolution and/or  hadronic wave function.

We note that much discussion recently has been in the context of so called "glasma" \cite{glasma}, which
refers to interacting system of produced gluons immediately after the collision. We feel that placing discussion squarely in this setup
 is not necessarily useful, and will start therefore by pointing out that  the  basic physics of correlations could be understood without
 reference to either "glasma" or "glasma flux tubes" .

The issue to some extent is in what frame one wishes to discuss the physics. The center of mass frame is not the simplest in the following sense. A very simple way of thinking about high energy scattering is the eikonal approximation, in which very energetic projectile partons scatter eikonally off the target fields. However if we are interested in particles produced at or close to central rapidity (like in the CMS data), we cannot use eikonal approximation in the center of mass frame. The incoming partons are stopped and loose most of their longitudinal momentum, thus clearly the eikonal approximation breaks down. To have a valid description of particle production in the center of mass frame one has therefore to take into account nonlinear interactions between the produced particles, at least on the level of classical approximation \cite{classical}.
The interacting system of gluons in the center of mass frame can then be viewed as "glasma".
 On the other hand if we follow the same exact process in the lab frame, it looks quite different. The incoming particles are indeed very energetic and they scatter by a very small angle with $p^+\gg p_T$. Thus recoil is negligible and eikonal approximation is applicable at high enough energy.
 Of course the physics is still the same. Thus for example even in the lab frame one has longitudinal electric and magnetic fields develop directly after collision, except that here their interpretation is rather less exotic than "glasma flux tubes"  \cite{eikonal}. 

The wave function of the incoming projectile carries a valence color charge density Lorentz contracted to a plane $J^-_a(x,x^-)=\rho^a(x)\delta(x^-)$. These charges create the Weizsacker-Williams field of softer gluons $b_i^a(x)$. The WW vector potential is determined through the classical equations of motion\cite{mv}
\begin{equation}\partial_i b^a_i(x) =\rho^a(x); \ \ \ \ \ \ 
\partial_ib^a_j-\partial_jb^a_i-gf^{abd}b^b_i(x)b^c_j(x)=0\label{ww}
\end{equation}
The first of these equations is essentially the Gauss' law, while the second defines $b$ as a two dimensionally pure gauge vector potential.
The color electric and color magnetic fields that correspond to this vector potential are localized in the plane containing the color charge density
$F^{+i}=b_i(x)\delta(x^-)$.
When this configuration of charges and fields passes through the target, it experiences a simple eikonal rotation
\begin{equation}
\rho(x)\rightarrow \bar\rho(x)=S(x)\rho(x); \ \ \ \ \ \ \ b_i^a\rightarrow\bar b_i^a=S^\dagger b^a_i[S\rho];\ \ \ \ \ \ \ \ F^{+i}=\bar b^a_i(x)\delta(x^-)\,.
\end{equation}
Here $S(x)$ is a unitary matrix determined by the target color fields, which rotates the color of the incoming parton wave function.
The effect of such a rotation is such that after the collision
\begin{equation}
\partial \bar b_i\ne \bar\rho
\end{equation}
Thus Gauss' law requires the presence of longitudinal electric field. In the light front the longitudinal field is not an independent degree of freedom, but rather is given by
\begin{equation}
E_3={1\over \partial^+}\left[D_iF^{+i}-\bar\rho\delta(x^-)\right]=\theta(x^-)\partial_i\Big[S^\dagger b_i[S\rho]-b_i[S\rho]\Big]\ne 0\,.
\end{equation}
The longitudinal magnetic field is also generated by the eikonal scattering, since the vector potential after scattering is not a two dimensional pure gauge anymore
\begin{equation}
F_{ij}=\Big[\partial_i\bar b^a_j-\partial_j\bar b^a_i-gf^{abd}\bar b^b_i(x)\bar b^c_j(x)\Big]\theta(x^-)\,.
\end{equation}
We stress again, the only physical process we are discussing is the eikonal scattering of gluons of the projectile on the static target. The origin of the longitudinal fields is therefore simply the Weizsacker - Williams fields of the gluons which are scattered out of the incoming beam. Once the direction of propagation of gluons is not parallel to the $z$ axis, the color electric and magnetic fields which are transverse to the direction of propagation have a nonvanishing projection onto the $z$ axis - hence  the appearance of the "longitudinal fields". This is a purely kinematical effect and it does not involve any dynamics of formation of flux tubes, or even interaction of produced gluons as such. In the rest of this paper we will therefore not refer  to glasma flux tubes, and our view of the scattering process will be that of a straightforward eikonal scattering of the projectile gluons on the color fields of the target.

\section{Where do correlations come from?}
We first discuss a very simple picture of the origin of correlations, and then confront it with the available today, albeit incomplete formulae for the double inclusive distribution.

Consider high energy scattering of a hadronic projectile on a stationary target in the lab frame. Since the projectile is very energetic, its wave function is approximately boost invariant. The boost invariance is of course only approximate, since at too high energy the rapidity evolution is important, and that introduces rapidity dependence inside the wave function. However for rapidity intervals $\Delta Y<{1\over\alpha_s}$ the evolution is not important\cite{glv}, and thus can be neglected if the produced particles are separated by rapidity interval which is not parametrically large.

The boost invariance leads naturally and straightforwardly to long range rapidity correlations. Simply put, the incoming wave function is the same at rapidity $Y_1$ and $Y_2$. The gluon distribution at rapidity $Y_1$ and $Y_2$ are the same, these gluons scatter exactly on the same target, and thus whatever happens at $Y_1$ also happens at $Y_2$. If for a particular target field configuration a gluon is likely to be produced at $Y_1$ at some impact parameter, a gluon is also likely to be produced at $Y_2$ at the same impact parameter: {\it e voila} - correlations.  This is especially true in the context of the projectile wave function dominated by the large "classical" Weizsacker-Williams field, since in this case fluctuations in the wave function are small and the gluon density {\it configuration by configuration} is almost the same at all rapidities .
This is the property of the hadronic wave function at high energy \cite{bubbles}.
\begin{equation}
|\Psi \rangle =\exp\Big\{i\int d^2xb^a_i(x)\int d\eta \left(a^{\dagger a}_i(x,\eta)+a^a_i(x,\eta)\right)\Big\}B(a,a^\dagger)|\psi\rangle \,.
\label{wf}
\end{equation}
Here $\psi$ is the wave function of valence charges, determining the distribution of the charge density $\rho$, $B$ is a Bogolyubov -type operator of the soft gluon fieds $a$, and the Weizsacker-Williams field $b$ is given in terms of $\rho$ by eq.(\ref{ww}). For large projectile the WW field is parametrically large $b\sim {1\over g}$, while the Bogolyubov operator $B$ produces the fluctuations of the gluon field of order unity. Thus for fixed $\rho(x)$ the gluon density fluctuates very weakly around large average value determined by the classical field
\begin{equation}
n=\langle a^\dagger a\rangle \propto b^2\sim O({1\over \alpha_s}); \ \ \ \ \ \ \ \ \langle n^2\rangle -\langle n\rangle ^2\sim 1
\end{equation}
The smallness of the fluctuations is clearly helpful. Although {\it the wave function} at different rapidities in a boost invariant projectile must be the same, the magnitude of the {\it color field} (and therefore the number of gluons) may differ at different values of $Y$ for the same configuration of the valence color charge density $\rho$, if the fluctuations in this wave function are significant. Thus in the same scattering event there may be significant differences between particle production at different rapidities. 
Still, although the quasiclassical nature of the state eq.(\ref{wf}) ensures long range rapidity correlations at large values of $\rho$, it is not absolutely necessary. Even in the presence of considerable fluctuations in the soft gluon wave function, one nevertheless would expect  positive correlations in rapidity. The only really necessary condition is that the density of incoming partons is large enough, so that there is large probability to produce more than one particle at a given impact parameter (we will quantify what we mean by "given impact parameter" shortly).

Thus the long range rapidity correlations come practically for free whenever the energy is high enough so that the wave function of the incoming hadron is approximately boost invariant, and there is very little in the actual dynamics of the collision that can affect this feature.  But by almost exactly the same logic we must conclude that positive angular correlations are also  almost unavoidable. Indeed, if two gluons hit the target at the same impact parameter, their scattering amplitude is determined by the same configuration of the target field. Thus if the  first gluon is likely to be scattered with momentum $\bf q$, the same is true for the second gluon. One therefore expects clear forward correlations for gluons that scatter at the same impact parameter. Of course, the two gluons will not scatter always with exactly the same momentum transfer even if they hit at exactly the same impact parameter, since even a fixed configuration of target fields corresponds to a nontrivial probability distribution of momentum transfer. Nevertheless, given that this distribution has a maximum at some particular momentum transfer, the angular correlations must be very generic.

The previous discussion is clearly oversimplified, since it does not address some important points. For example, for a soft gluon to be produced in the final state, it is not enough for it to acquire some transverse momentum. It also must decorrelate from the valence charge that emitted it in the incoming wave function. Otherwise it will not be produced as a particle in the final state, but rather as part of the Weizsacker-Williams field of the produced valence parton. We will therefore turn to an explicit formula that determines the gluon double inclusive spectrum in order to see to what extent this explicit expression is consistent with our simple discussion.

Calculation of multigluon amplitudes at high energy has been a subject of several papers in recent years \cite{jamal},\cite{baier},\cite{multigluons},\cite{francois},\cite{pomloops2}. Nevertheless, unfortunately we still do not have a final formulae which can be used to analyse p-p collision at high energy. The approach of \cite{francois} is suited for applications when both colliding hadrons are dense, while that of \cite{jamal},\cite{baier},\cite{multigluons} is appropriate when one of the objects is dense and one dilute. The actual situation in p-p collisions at LHC is probably one where the density in the proton wave function is still not parametrically large, but is already not perturbatively small. In that sense the expressions of \cite{baier} are probably  more appropriate, since they do not discard terms which are leading at low densities, which is done in \cite{francois}. At any rate, we believe that qualitative features of our discussion are general enough, and we will use the explicit formulae from \cite{baier}. 

According to \cite{baier} (see also \cite{multigluons}) the inclusive two gluon production probability is given by
\begin{equation}
{dN\over d^2pd^2kd\eta d\xi}=\langle A_{ij}^{ab}(k,p)A_{ij}^{*ab}(k,p)\rangle _{P,T}
\label{section}
\end{equation}
with the amplitude
\begin{eqnarray}
A_{ij}^{ab}(k,p)&=&
 \int_{u,z}e^{ikz+ipu}\int_{x_1, x_2} 
     \Big\{f_i( z -  x_1)
     \left[S( x_1)-S( z)\right]^{ac}\rho^c(x_1)\Big\}\Big\{f_j(u - x_2)\left[ S( u)- S(x_2)\right]^{bd}\rho^d(x_2)\Big\}\nonumber\\
&&-{g\over 2} \int_{x_1} 
f_i(z -  x_1)
  f_j(u-  x_1)
  \Big\{\left[S( x_1)-S( z)\right]\tilde\rho( x_1)
  \left[ S^\dagger(u)+S^\dagger(x_1)\right]\Big\}^{ab}\nonumber\\
&& + g\int_{ x_1} 
f_i( z - u)
 f_j( u - x_1)
     \left\{ \left( S( z) - S( u) \right)
        \tilde\rho( x_1) S^{\dagger}( u)\right\}^{ab} \, .\label{amplitude}
\end{eqnarray}
Here 
\begin{equation} f_i(x-y)={(x-y)_i\over (x-y)^2}\end{equation}
and we have defined $\tilde\rho\equiv -i T^a\rho^a$. The charge density is normalized such that for a single gluon $\rho^a=gT^a$.
In these formulae $\rho^a(x)$ is the valence color charge density in the projectile wave function, while $S^{ab}(x)$ is the eikonal scattering matrix determined by the target color fields.
The average in eq.(\ref{section}) denotes averaging over the projectile and the target wave functions. We also note that in this expression the gluon with momentum $p$ is assumed to have larger rapidity, and thus the emission of the two gluons is not completely symmetric.

The actual equation given in \cite{baier} is slightly different and this requires explanation. The color charge density appearing in \cite{baier} are quantum operators with quantum commutation relations of $SU(N)$ algebra. On the other hand $\rho^a$ in eq.(\ref{amplitude}) are $c$-number functions, and averaging over $\rho$ is understood as averaging over a classical ensemble 
\begin{equation}\langle O\rangle _P\,=\,\int D\rho \,W_P[\rho] \,O\ .\label{average}\end{equation}
 As discussed at length in \cite{kl}, the full quantum averaging is equivalent to classical averaging procedure eq.(\ref{average}) for totally symmetrized products of operators $\hat \rho^a$. To obtain eq.(\ref{amplitude}) from the expression given in \cite{baier} we have rewritten the original expression of \cite{baier} in terms of the anticommutator of two $\hat \rho$'s in the first line. Thus the "classical" expression $\rho^a\rho^b$ is equivalent to the quantum operator ${1\over 2}\{\hat \rho^a,\hat \rho^b\}$. This procedure of reordering generates an additional term in eq.(\ref{amplitude})  which is not present in \cite{baier}.
 
 The physical meaning of the three terms in eq.(\ref{amplitude}) is straightforward. The first term corresponds to independent production of the two gluons. This term is leading in the limit of large color  density $\rho\sim 1/g$. One should keep in mind, however that in this limit other terms not included in eq.(\ref{amplitude}) are equally important \cite{francois},\cite{pomloops2}. The second term corresponds to production of two gluons emitted from the same color source in the incoming projectile wave-function. The third term corresponds to the process whereby the softer gluon has been emitted in the wave function by the harder one, with both gluons subsequently produced in the collision. In terms of BFKL ladders, the (square of the) first term is a part of the diagram containing two independent ladders, while the (square of the) last two terms describe emission of two gluons contained in the same BFKL ladder.
 
To calculate the cross section one has to square the amplitude. This produces many terms, since in accordance with our previous discussion one has to symmetrize the factors of charge density between the amplitude and the conjugate amplitude. The full expression for the cross section is given in the Appendix A. Here in the text we only reproduce one part of this expression which arises from squaring the first term in the amplitude  eq.(\ref{amplitude}) which is responsible for independent production of the two gluons.
\begin{equation}
{dN\over d^2pd^2kd\eta d\xi}=\langle\sigma^4+\sigma^2+\sigma^3\rangle _{P,T}\label{production}
\end{equation}
with
\begin{eqnarray}
\sigma^4&=&\int_{u,z,\bar u,\bar z}e^{ik(z-\bar z)+ip(u-\bar u)}\int_{x_1, x_2,\bar x_1,\bar x_2} \vec{f}( \bar z - \bar x_1)\cdot \vec{f}( x_1- z)\, 
             \vec{f}({\bar  u}-\bar x_2)\cdot \vec{f}( x_2- u) \label{sigma4}\\
             &&     \times
              \left\{ \rho(x_1)[S^\dagger( x_1)-S^\dagger( z)]
                       [S(\bar x_1)-S(\bar z)]\rho(\bar x_1)\right\} \left\{\rho(x_2)
               [S^\dagger(u)-S^\dagger(x_2)][S({\bar u})-S(\bar x_2)\rho(\bar x_2)\right\}  \nonumber
               \end{eqnarray}
  
The explicit expressions for $\sigma^2$  and $\sigma^3$ are given in the Appendix A. Although these are long expressions, only the term given in eq.(\ref{sigma4}) is relevant for most of our discussion.  This is precisely the contribution which has the physics discussed above. The two gluons here are produced independently from each other, but from exactly the same configuration of sources through scattering on the same target field. In the other terms the two gluons are correlated with each other in the incoming wave function, and thus these terms contain additional physics. Let us therefore consider the cross section of eq.(\ref{sigma4}). It is very easy to see that it indeed produces angular correlations. One can write it as
\begin{equation}
\sigma^4(k,p)=\langle\sigma(k)\sigma(p)\rangle _{P,T}\label{sigma4again}
\end{equation}         
where
\begin{equation}
\sigma(k)=\int_{z,\bar z}e^{ik(z-\bar z)}\int_{x_1,\bar x_1} \vec{f}( \bar z - \bar x_1)\cdot \vec{f}( x_1- z)\, 
                          \left\{ \rho(x_1)[S^\dagger( x_1)-S^\dagger( z)]
                       [S(\bar x_1)-S(\bar z)]\rho(\bar x_1)\right\} \,.
                        \label{sigma}
\end{equation}           
For {\it fixed configuration} of the projectile sources $\rho(x)$ and target fields $S(x)$, the function $\sigma(k)$ as a function of momentum has a maximum at some value ${\bf k}={\bf q}$.
Therefore clearly the product in eq.(\ref{sigma4again}) is maximal for ${\bf k}={\bf p}={\bf q}$. The value of the vector ${\bf q}$ of course differs from one configuration to another, but the fact that momenta $\bf k$ and $\bf p$ are parallel does not. Therefore after averaging over the ensemble $\sigma^4(k,p)$ has maximum at relative zero angle between the two momenta.

 We reiterate, that even though averaged over all configuration $\langle\sigma(k)\rangle _{P,T}$ must be isotropic, there is absolutely no reason for it to be isotropic for any given configuration. The strength of the maximum of course depends on the detailed nature of the field configurations constituting the two ensembles (the projectile and the target). We will discuss some qualitative features of these in the next section. But first, it is interesting to ask is the maximum of $\sigma(k)$ unique, or perhaps there is finite degeneracy. It is in fact easy to see that the maximum is doubly degenerate. The probability $\sigma(k)$ can be written in terms of the single gluon production amplitude $a(k)$
\begin{equation}
a^a_i(k)=\int dz e^{ikz}\int_{x_1} f_i(  z -  x_1) [S( x_1)-S( z)]^{ab}\rho^b( x_1)\,. \label{singleamp}
\end{equation}
Since the amplitude $a$ is real in {\it coordinate space}, we have
\begin{equation}            
 \sigma(k)=a(k)a^*(k)=a(k)a(-k)\,.
 \end{equation}
Configuration by configuration this is clearly symmetric
\begin{equation}
\sigma(k)=\sigma(-k)\,.
\end{equation}           
The "classical"' contribution to the two particle inclusive production probability is therefore symmetric 
under
\begin{equation}
\sigma^4(k,p)=\sigma^4(-k,p)\,. \label{symmetry}
\end{equation}
and must have two degenerate maxima - at relative angles $\Delta\phi=0,\pi$.      
The feature of the equal strength of forward and backward correlations stays true beyond the specific form of the single gluon production amplitude  eq.(\ref{singleamp}).  In the limit of dense projectile eq.(\ref{sigma}) is not strictly applicable. However, although one does not have an analytic formula valid in this limit, the formalism of \cite{francois} expresses the double gluon production probability in terms of a solution of classical equations of motion. This solution, which is precisely the single gluon production amplitude, is real in coordinate space for any fixed configuration of projectile and target fields. The reality of the amplitude in the coordinate space is the only property required to establish degeneracy, and thus degeneracy persists in the dense projectile limit as well. We note that numerical results of \cite{tuomas} do not seem to display the exact symmetry of eq.(\ref{symmetry}), which may be an indication of some subtlety of the numerical procedure of \cite{tuomas}. 

The other terms in the gluon production amplitude eq.(\ref{amplitude}) are not likely to lead to correlations of the type just described. 

The third term in eq.(\ref{amplitude}), where the gluon produced at the point $z$  is emitted from the other observed gluon at the point $u$, disfavors production at the same impact parameter because of the suppression factor $S(u)-S(z)$. The two gluons when scattered at the same impact parameter do not decohere, but rather scatter as a single coherent state, with the gluon at $z$ emerging in the final state as part of the Weizsacker-Williams field of the gluon at $u$. On the other hand whenever the two gluons do decohere, since they were correlated in the incoming wave function, they emerge in the final state with large relative transverse momentum. Thus this particular term in the amplitude mostly leads to back to back production in the final state and is responsible for the large away side, rapidity independent maximum at relative angle $\pi$, prominently present in the data. 

The second term in eq.(\ref{amplitude}) favors production at the point $u$ close to $x_1$, but $z$ far from $x_1$. Thus one expects the momentum of the gluon produced at $z$ to be uncorrelated with that of the gluon produced at $u$. Whenever the gluon at $u$ is produced with significant transverse momentum, the balancing transverse momentum resides at the "valence" rapidity. This term is therefore responsible for the away side peak between one of the observed particles and another particle produced at a more forward rapidity.

One can estimate the overall magnitude of the correlation by the following simple argument. In order for two produced gluons to be correlated in the final state, they have to be close in the initial state and also scatter off the same target field. We will assume that both the target and the projectile are characterized by corresponding saturation momenta $Q_s^{P(T)}$. The inverse of the correlation momentum is the correlation length in the hadron $L\sim 1/Q_s$. It is reasonable to expect that typical field configurations contributing to the hadronic ensemble of, say the target, have variation only on distance scale greater than $1/Q_s$. Thus the two gluons that hit the target at distance $x<1/Q_s^T$ apart from each other scatter on the same field. By the same argument, for the two incoming gluons to be in the same state they have to be located in the impact parameter plane no further than $1/Q_s^P$ away from each other.
 Thus for correlated production the two gluons need to be within the radius ${1\over Q_s^{max}}$ of each other, where $Q_c^{max}$ is the larger of the two saturation momenta $Q_s^P$ and $Q_s^T$. On the other hand the total number of produced gluons is proportional to the total transverse area of the smaller between the two objects participating in collision. Thus parametrically
\begin{equation}
\left[{d^2N\over d^2pd^2k}-{dN\over d^2k}{dN\over d^2p}\right]/{dN\over d^2k}{dN\over d^2p}\sim    {1\over (Q_s^{max})^2S_{min}}\,.
\label{estimate}
\end{equation}
This estimate is of course parametrically the same as given in \cite{glv}.

We would like at this point to make contact with the recent paper \cite{ddgjlv}. The calculation of gluon production in \cite{ddgjlv} is based on simplified version of eq.(\ref{sigma4again}) supplemented with specific prescription for averaging over the projectile and target fields. Specifically, \cite{ddgjlv} expands the scattering matrix $S$ to first order in target fields, and keeps only the leading term $S(x)\rightarrow 1+\alpha(x)$. The expression for $\sigma^4$ then becomes a homogeneous function of the target and projectile fields
\begin{equation}
\sigma^4\sim (\rho\alpha\alpha\rho)(\rho\alpha\alpha\rho)\label{expand}
\end{equation} 
For simplicity we suppress the color indices and transverse coordinates on all the functions. One next averages over the charge densities assuming Gaussian ensemble
\begin{equation}
\langle\rho^4\rangle =3\langle\rho^2\rangle \langle\rho^2\rangle 
\end{equation} 
and similarly for the target. And finally the high energy evolution is included by substitution
\begin{equation}
\langle\rho(x)\rho(y)\rangle \rightarrow \Phi(x-y)
\end{equation}
with $\Phi$ taken to be a solution of the Balitsky-Kovchegov equation \cite{bk}. Although the angular distribution has not actually been calculated in \cite{ddgjlv}, the authors argued that the correlation should in fact have a maximum at collinear momenta. The subsequent numerical investigations based on the same approach bear this out \cite{kevin}. 

Our general discussion provides an intuitive explanation for this result and also makes it clear that the presence of the correlations does not depend on the specifics of the approximation used to estimate ${d^2N\over d^2kd^2p}$. The magnitude of the effect however may depend on the approximation quite strongly. We next want to comment on this issue.

\section{Issues with averaging.}
 From eqs.(\ref{sigma4again},\ref{sigma}) we know that the basic  averages that one needs to calculate are of the type
\begin{equation}
\langle[S^\dagger(x)S(z)]^{ab}[S^\dagger(y)S(u)]^{cd}\rangle _T\label{tav}
\end{equation}
and similarly for the projectile
\begin{equation}
\langle\rho^a(x)\rho^b(\bar x)\rho^c(y)\rho^d(\bar y)\rangle _P\,.
\label{pav}
\end{equation}
The Gaussian averaging procedure described above is fairly restrictive, in the sense that as any Gaussian averaging it probably tends to underestimate correlations. 
In particular Gaussian averaging over color singlet ensemble necessarily puts the densities in eq.(\ref{pav}) pairwise into color singlet states. As pointed out in \cite{adrianjamal} this leaves out some possible configurations which are overall color singlets, but where no two factors of $\rho$ form a color singlet separately. This for example happens, when the factors of $\rho$ are pairwise in color octets, with the two octets forming an overall singlet. Such configurations in principle can also contribute to the correlated part of the particle production. Formally they are suppressed in the large $N_c$ limit. However the correlated part of the production probability itself {\it when calculated with the Gaussian averaging} is also suppressed by $1/N_c^2$ relative to the uncorrelated part, and thus omission of these terms may be dangerous \cite{adrianjamal}. Physically these terms correspond to interference contributions. For example when the two factors of $\rho$ in eq.(\ref{sigma}) are in an octet, this corresponds to a situation when the charge densities in the amplitude and complex conjugate amplitude are different, but still the same gluon in the final state is produced due to the difference in the scattering factors $S$ in the amplitude and the conjugate amplitude. 

Although these $1/N_c^2$ suppressed terms are interesting, taking them properly into account requires one to go beyond the dipole model \cite{dipole} and the BK equation, and in the dense region studying the full B-JIMWLK evolution \cite{jimwlk}.
However it is not obvious that even in the leading order in $1/N_c$ the Gaussian approximation is adequate to discuss correlated production. 
Here we would like to discuss only these leading order terms. We will argue that Gaussian averaging procedure is likely to miss terms in the correlated production probability which are of the same order in $1/N_c$ as the uncorrelated piece. 

The leading $N_c$ piece in eq.(\ref{sigma4again}) comes from the configuration where the charge densities in each one of the single gluon production amplitudes are in the color singlet. The relevant average to calculate is
\begin{equation}
\langle \rho^a(x_1)\rho^a(\bar x_1)\rho^b(x_2)\rho^b(\bar x_2)\rangle_P
\langle{\rm Tr}\left\{[S^\dagger( x_1)-S^\dagger( z)]
                       [S(\bar x_1)-S(\bar z)]\right\}{\rm Tr}\left\{[S^\dagger( x_2)-S^\dagger( u)]
                       [S(\bar x_2)-S(\bar u)]\right\}\rangle_T\,.
                       \label{avera}
\end{equation}
Let us first concentrate on the projectile average. As mentioned above, averaging with a Gaussian weight one obtains in the leading order in $1/N_c$
\begin{equation}
\langle \rho^a(x_1)\rho^a(\bar x_1)\rho^b(x_2)\rho^b(\bar x_2)\rangle_{Gaussian}=\langle \rho^a(x_1)\rho^a(\bar x_1)\rangle_{Gaussian}\langle\rho^b(x_2)\rho^b(\bar x_2)\rangle_{Gaussian}\,.
\label{factorization}
\end{equation}
In this approximation therefore, clearly the correlated piece in the production probability vanishes, and only the subleading in $1/N_c$ correction resurrects the correlations. We stress however, that this is not the result of the leading $N_c$ approximation {\it per se}, but rather of the Gaussian averaging procedure.

It may be tempting to think that factorization in the large $N_c$ limit is natural due to presence of large number of degrees of freedom, and therefore in some sense large $N_c$ might act similarly to heavy nucleus. However this is not the case. Even though the number of degrees of freedom is large, even in the large $N_c$ limit the theory has legitimate states which contain small number of particles. A color dipole is an example of such state. It is a superposition of many states (different color orientations) of two particles, rather than a state with many particles. In a state like this the central limit theorem does not hold, the fluctuations in density can be large even in the large $N_c$ limit, and it is the large fluctuations in the ensemble that break factorization of correlation functions.

In fact a very similar question was considered a while ago in \cite{yoshi} in connection with factorization of dipole densities in the dipole model\cite{dipole}. Indeed the observable we are interested in eq.(\ref{avera}) is rather similar to the dipole density
\begin{equation}
n(x_1,\bar x_1)=\Big(\rho^a(x_1)-\rho^a(\bar x_1)\Big)^2\,.
\end{equation}
As shown in \cite{yoshi} within the dipole model (which is defined entirely within the  large $N_c$ limit\cite{kl1}) the product of two densities does not factorize, but rather behaves as
\begin{equation}
\langle n(x_1,\bar x_1)n(x_2,\bar x_2)\rangle- \langle n(x_1,\bar x_2)\rangle \langle n(x_2,\bar x_2)\rangle \sim
 \langle n(x_1,\bar x_2)\rangle \langle n(x_2,\bar x_2)\rangle b^{-\lambda}\label{corden}
 \end{equation}
 where $b$ is the transverse distance between the two dipoles and $\lambda$ is a number, whose exact value is unimportant for us. This result hods in the limit where the distance between the two dipoles is much greater than their respective sizes, and thus it does not display any angular correlation between the orientations of the two dipoles.   Nevertheless eq.(\ref{corden})  clearly exhibits the fact that factorization is not an inherent property of the large $N_c$ limit. 
 Once we accept that the factorization is broken, it is natural to expect that the actual correlation function in the regime where the two dipoles overlap in space, also exhibits angular correlations in the orientation of the two dipoles. 
 
 Note that this is precisely the regime relevant to our discussion of angular correlations in emission. The {\it same} configuration of color charges produces the same gluons (at different rapidities), which produce correlated hadrons in the final state. Thus the most important region of the phase space is when all four points in the correlator eq.(\ref{avera}) are close to each other, in the sense that they are all within the correlation length $1/Q_s$. It is very hard to imagine that in this regime factorization holds (see \cite{LL,klw} for more discussion of such correlations). 
Thus we indeed expect that any realistic non-Gaussian weight function for the ensemble averaging will lead to a nonvanishing contribution to the correlated piece of gluon production even in the large $N_c$ limit.
 
Turning to the target averaging in eq.(\ref{avera}), the terms that have to be averaged are of the type of observables described in the large $N_c$ limit by the dipole model \cite{dipole} 
\begin{equation}
\langle{\rm Tr}\left\{[S^\dagger( x)S( z)]
                       \right\}{\rm Tr}\left\{[S^\dagger( y)S( u)]
                       \right\}\rangle_T=\langle s(x,z)s(z,x)s(y,u)s(u,y)\rangle_T\label{fad}
\end{equation} 
 where $s(x,y)={\rm Tr}[S^\dagger_F(x)S_F(y)]$ - is the  scattering amplitude of the fundamental dipole, and the equality in eq.(\ref{fad}) holds in the large $N_c$ limit.
 The approximation which is frequently used in the literature to calculate the averages of this type also invokes factorization
 \begin{equation}
 \langle s(x,y)s(u,v)\rangle=\langle s(x,y)\rangle\langle s(u,v)\rangle\,.
 \label{fact}
 \end{equation}
 The target averaging of \cite{ddgjlv} would follow from this approximation in the limit of weak fields. 
 When the dipole $(x,y)$ is far from the dipole $(u,v)$ (much further than $1/Q_s$), the factorization is a good approximation since the fields on which the scattering amplitude is calculated are not correlated with each other. However, as before we are clearly interested in the case where all the points are within the distance of order $1/Q_s$ or smaller. In this case, just like for the projectile, the factorization of eq.(\ref{fact}) is not a property of the large $N_c$ limit but is rather an {\it ad hoc} assumption, used only due to its simplicity. 
 
 Strict factorization of the type eq.(\ref{fact}) is only possible if the statistical ensemble consists of a single configuration. There is however no reason to expect that in the large $N_c$ limit fluctuations around some leading configurations are suppressed by powers of $1/N_c$.
For example, the energy evolution of $s$ is given by the dipole model Hamiltonian, which does not contain $N_c$ at all. The probability distribution of the dipole model $W[s]$ evolves with rapidity according to \cite{LL,kl1}
\begin{equation}
 {d\over d Y}W[s]=
 \frac{ \bar{\alpha}_s}{2\,\pi}\,
\int_{x,y,z}\frac{(x-y)^2}{(x-z)^2\,(z\,-\,y)^2}\,\left[\,s(x,\,y)\,-\,
\,s(x, z)\,s(y,z)\,\,\right]
\frac{\delta}{\delta s(x, y)} W[s]
\end{equation}
with $\bar \alpha$ - the 'tHooft coupling, which is finite at infinite $N_c$.
Thus any nontrivial initial distribution $W[s]$ evolves smoothly to higher energy and remains nontrivial.

In fact it would be very interesting to see what happens to correlations that one can put into the initial ensemble, as rapidity grows. Technically one should choose an ensemble $W_0[s]$ of initial configurations $s(x,y)$,  which contains short range correlations. These correlations should be confined to within the saturation radius, that is $\langle s(x,y)s(u,v)\rangle \ne \langle s(x,y)\rangle\langle s(u,v)\rangle$ only when $|x-u|, |y-v|<1/Q_s$.
 Each configuration of the ensemble should be evolved independently according to the BK equation \cite{kl1}. The correlations at the final rapidity are then calculated by averaging the correlator calculated in the final ensemble over the ensemble of initial conditions. 
 
 This procedure is the same as the one implemented in \cite{nestor}. However in order to study the angular correlations between the produced particles, it is important to understand angular correlations in the target ensemble. Therefore the individual members $s(x-y)$ of the initial ensemble should not be isotropic. The rotational invariance must be restored by averaging over the whole ensemble rather than configuration by configuration.
 
As the rapidity grows, so does $Q_s$ and the correlation radius shrinks so  that one naively expects only the points inside this new saturation radius $1/Q_s(Y)$ to stay correlated. We note however that it is also possible that the correlation length in this particular channel is larger than $1/Q_s$. For example it was shown recently in \cite{munier} in a QCD like model, that the saturation momentum itself varies rather slowly in the impact parameter plane. In some sense one can ascribe to it a correlation length which is far greater than $1/Q_s$, and is instead $\lambda \propto {1\over Q_s}e^{c\ln^21/\alpha_s}$. Although there is no particular reason to expect that such a large correlation length will also determine the scale of disappearance of correlations of the dipole amplitudes, if it does it would be very interesting and one would have to revise the estimate eq.(\ref{estimate}) accordingly. 
 
 To summarize, there are good reasons to expect that the factorization of both projectile and target averages is broken {\it at leading order in $1/N_c$} in the kinematical domain relevant to the correlated production of particles. To study this question one certainly has to go beyond simple rotationally invariant solutions of the BK equation. While technically challenging, it would be very interesting to understand and quantify this effect.

\section{Conclusions}
We conclude with a short summary. We have shown that rapidity and angular correlations are a very general feature of particle production at high energy. They are an automatic consequence of the boost invariance of the projectile wave function, provided two conditions are met. 

First, there should be some "classical" component in the (projectile) wave function, meaning that the short range rapidity fluctuations in the wave function should not be overwhelming and should not lead to complete decorrelation between gluons at different rapidities. This condition is satisfied in QCD for dense as well as dilute projectiles. In the dense case the soft gluon wave function is dominated by the "classical" rapidity independent Weizsacker-Williams field. For dilute projectile fluctuations are significant, but their contribution to gluon density is not parametrically larger than that of the WW field. 

Second, the particle density/scattering probability must be large enough so that there is non-negligible probability to produce more than one particle at fixed impact parameter within the saturation radius $1/Q_s$. This condition is satisfied when both colliding objects are dense and the values of the two saturation momenta are not vastly different. In that case the number of gluons in the incoming wave function within area $1/Q^2_s$ is of order $1/\alpha_s(Q_s)$, while the probability of producing one particle is of order one. For scattering of two dilute objects this condition is not met. However at high enough energies the density in the wave function grows and the effect must become visible.

The angular correlations originate from configuration by configuration fluctuations of the projectile and target structure in the transverse plane away from a rotationally invariant state. 
The effect we have discussed here has several tell-tale features. Produced particles are correlated in angle, with forward and backward correlations being of equal strength in the case where the two colliding objects are nuclei. When the colliding objects are not dense, there is an additional contribution to particle production, from a "single ladder" which  significantly enhances back to back correlations. This contribution is responsible for the bulk of the observed back to back correlated production. The correlation is present also in the magnitude of the transverse momentum and not just in the angle. The latter correlation in fact does not require local rotational asymmetry of the projectile/target configurations. It would be interesting to try and measure these correlations as well. 

The relative magnitude of the forward correlations should initially increase with energy for p-p collisions, since the relative importance of the single ladder terms diminishes.  
Interestingly however, the estimate of eq.(\ref{estimate}) suggests that the effect decreases with energy, once the colliding systems can be treated as saturated objects with well defined saturation momentum, since the saturation momentum grows with energy. Thus at very high energies the effect should disappear. If we apply this logic also to nuclear collisions, we should conclude that the effect if observed by  ALICE should be significantly smaller than that observed by PHOBOS and STAR. Our discussion of course disregards the effects of flow, which are generally believed to be very important for nucleus-nucleus collisions. The latest STAR data \cite{star} support this view. It is possible therefore that our considerations about angular correlations are not valid for nuclear collisions, in the sense that the main mechanism of collimation is indeed due to the flow. It would nevertheless be interesting to try and disentangle the flow effects from the intrinsic correlations in the initial state discussed in this note. We also note that the estimate eq.(\ref{estimate}) refers not only to angular correlation, but rapidity correlation in general. Thus independently of the question of radial flow, if the observed long range rapidity correlations are due production from correlated domains in the boost invariant incoming wave function, the trend should be that of decreasing correlated production going to higher energy.

Finally, we have argued that the correlations must survive also in the leading order in $1/N_c$ expansion. Their subleading nature in current numerical implementations is due to the factorization assumption which is not valid in the region of the phase space relevant for the correlated production. We believe that improvement of this aspect of current calculations is imperative in order for the results to be quantitatively reliable.

\appendix

\section{Ordering}
In this Appendix we derive the completely symmetrized expression for the doubly inclusive gluon production. When averaging this expression over the projectile wave function, the color charge density $\rho$ should be treated as classical commuting variable. Simply squaring the expression for the amplitude in eq.(\ref{amplitude}), one gets eq(\ref{production}) with $\sigma^4$ given in eq.(\ref{sigma4}) and
\begin{eqnarray}
  \sigma^2&=&\int_{u,z,\bar u,\bar z}e^{ik(z-\bar z)+ip(u-\bar u)}\int_{x_1,\bar x_1}\\
  &&{g^2\over 4}\vec{f}({\bar  z} - \bar  x_1)\cdot \vec{f}( x_1- z)\, 
             \vec{f}({\bar u}-\bar  x_1)\cdot \vec{f}( x_1- u)\nonumber       \\  
       &&\times{\rm Tr}\Big\{ \left[ S( u)+S( x_1)\right]\bar \rho(x_1) \left[S^\dagger(x_1)-S^\dagger(z)\right]\left[S(\bar x_1)-S(\bar z)\right]\tilde\rho(\bar x_1)\left[S^\dagger(\bar u)+S^\dagger(\bar  x_1)\right]\Big\}\nonumber\\
&&+          g^2  \vec{f}({\bar  z} - \bar u)\cdot \vec{f}( u- z)\, 
             \vec{f}({\bar u}-\bar x_1)\cdot \vec{f}( x_1- u)\nonumber\\
&&\times    {\rm Tr}\Big[  \left\{  S({ u})
        \tilde\rho( x_1)\left( S^\dagger( z) - S^\dagger( u) \right)\right\} 
         \left\{ \left( S({\bar  z}) - S({\bar u}) \right)
        \tilde\rho(\bar{ x}_1) S^{\dagger}({\bar u})\right\}     \Big]\nonumber\\
&&-{g^2\over 2}\vec{f}({\bar  z} - \bar{ x_1})\cdot \vec{f}(u- z)\, 
             \vec{f}({\bar  u}-\bar x_1)\cdot \vec{f}( x_1- u)\nonumber\\
&&\times      {\rm Tr}\Big[  \left\{ S( u)
        \tilde\rho( x_1)\left( S^\dagger({  z}) - S^\dagger( u) \right)\right\} \left[S(\bar x_1)-S({\bar  z})\right]\tilde\rho(\bar x_1)\left[ S^\dagger({\bar  u})+S^\dagger({\bar x_1})\right]\Big]\nonumber\\
&&-{g^2\over 2}\vec{f}({\bar  z} - \bar{ u})\cdot \vec{f}(x_1- z)\, 
             \vec{f}({\bar  u}-\bar  x_1)\cdot \vec{f}( x_1- u)\nonumber\\
&&\times   {\rm Tr}\Big[ \left[ S({ u})+S({ x_1})\right]\bar \rho(x_1)\left[ S^\dagger({x_1})+S^\dagger({ z})\right]
                      \left\{ \left( S({\bar  z}) - S({\bar  u}) \right)
        \tilde\rho(\bar x_1) S^{\dagger}({\bar  u})\right\}     \Big]\nonumber
\end{eqnarray}
\begin{eqnarray}
\sigma^3&=&\int_{u,z,\bar u,\bar z}e^{ik(z-\bar z)+ip(u-\bar u)}\int_{x_1,\bar x_1, x_2 (\bar x_2)}\\
&&  -{g\over 2}\vec{f}({\bar  z} - \bar x_1)\cdot \vec{f}(x_1- z)\, 
             \vec{f}({\bar u}-\bar x_1)\cdot \vec{f}( x_2- u)\nonumber\\
&&\times \rho(x_1)[S^\dagger( x_1)-S^\dagger( z)]
             \left[S(\bar  x_1)-S({\bar  z})\right]\tilde\rho(\bar x_1)\left[ S^\dagger({\bar  u})+S^\dagger({\bar x_1})\right]
            [S( u)-S(x_2)]\rho(x_2)\nonumber\\
&&-{g\over 2} \vec{f}({\bar  z} - \bar x_1)\cdot \vec{f}( x_1- z)\, 
             \vec{f}({\bar  u}-\bar x_2)\cdot \vec{f}( x_1- u)\nonumber\\
&&\times  \rho(\bar x_2)[S^\dagger(\bar u)-S^\dagger( \bar x_2)]
             \left[ S( u)+S( x_1)\right]\tilde\rho( x_1)\left[S^\dagger( x_1)-S^\dagger(z)\right]
            [S(\bar x_1)-S(\bar z)]\rho(\bar x_1)\nonumber\\
&&+g\vec{f}({\bar  z} - \bar u)\cdot \vec{f}( x_1- z)\, 
             \vec{f}({\bar  u}-\bar x_1)\cdot \vec{f}( x_2- u)\nonumber\\
             &&\times \rho(x_1)[S^\dagger( x_1)-S^\dagger( z)]
              \left\{ \left( S({\bar z}) - S({\bar  u}) \right)
        \tilde\rho(\bar x_1) S^{\dagger}({\bar  u})\right\}                          
            [S(u)-S(x_2)]\rho(x_2)\nonumber\\
              &&+g \vec{f}({\bar z} - \bar x_1)\cdot \vec{f}(x_1- z)\, 
             \vec{f}({\bar u}-\bar x_2)\cdot \vec{f}( x_1- u)\nonumber\\
&&\times \rho(\bar x_2)[S^\dagger(\bar u)-S^\dagger( \bar x_2)]
            S({  u})\tilde\rho( x_1)\left[S^\dagger( z)-S^\dagger({ u})\right]
            [S(\bar x_1)-S(\bar z)]\rho(\bar x_1)\,.\nonumber
\end{eqnarray}
However in this expression the factors of $\rho$ coming from the amplitude are not symmetrized with respect to factors coming from the conjugate amplitude. To derive the fully symmetric expression we instead follow the formalism of \cite{multigluons}.

The amplitude of a single gluon production in the dense-dilute scattering is given by
\beq\label{qjim}
Q^a_i(z)\,=\,\int\, d^2x f_i(z-x)\,[S^{ab}(z)\,-\,S^{ab}(x)]\, J^b_R[S,x]\,,\ \ \ \ \ \ \ \ \ \ \ \ \ \ \ \ \ 
J_R^a(S,x)=-{\rm tr} \left\{S(x)T^{a}{\delta\over \delta S^\dagger(x)}\right\}
\eeq
In this expression the projectile wave function has been "integrated out"\cite{multigluons}. It can be however "integrated back in" by representing the right rotation operatorsd as the projectile charge density operators acting on the projectile wave function\cite{kl1}:
\begin{equation}
J_R^a(x)|P\rangle\,=\,\hat\rho^a_x|P\rangle\,,\ \ \ \ \ \ \ \ \ \ \ \ \ 
J_R^a(x)J_R^b(y)|P\rangle\,=\,\hat\rho^b_y\hat\rho^a_x|P\rangle
\label{jays}
\end{equation}
The operator associated with inclusive production of a single gluon with momentum $k$ is
\beq\label{Og2}
{\cal O}_g(k)\,=\,
\int {d^2 z\over 2\,\pi}\,{d^2 \bar z\over 2\,\pi}\, \,e^{i\,k\,(z\,-\,\bar z)}\,
\,Q^a_i(z,[S])\,Q^a_i(\bar z,[\bar S])
\eeq
where $S$ is the single gluon S-matrix in the amplitude and $\bar S$ in the conjugate amplitude.
Double gluon production without rapidity evolution between the two produced gluons  is given by 
\begin{equation}
{dN\over d^2pd^2kd\eta d\xi}=\langle {\cal O}_g(k)\,{\cal O}_g(p)\rangle_{P,T}|_{S=\bar S}
\end{equation}
The operators $Q^a_i(z,[S])$ do not commute with each other. We will move all the right rotation operators $J_R$
to the right  and then convert them into the operators $\hat \rho$ according to eq.(\ref{jays}).  We also keep all operators  $J_R[\bar S]$ to the right of $J_R[S]$. 

The amplitude $A_{ij}^{ab}$ reads
\begin{eqnarray}
\label{A}
A_{ij}^{ab}(k,p)&=&\int_{z,u} e^{ikz+ipu} Q^a_i(z)Q^b_j(u)=\int_{z,u} e^{ikz+ipu}\int_{{x_1}{x_2}}\left[
 f_i(z-{x_1})[S_z-S_{x_1}]^{ac} f_j(u-{x_2})[S_u-S_{{x_2}}]^{bd}J^c_R({x_1})J^d_R({x_2})\right. \nonumber \\
 &+& f_i(z-{x_1})[S_z-S_{x_1}]^{ac}\,  f_j(u-{x_2})[S_u\delta_{u{x_1}}-S_{{x_2}}\delta_{{x_2}{x_1}}]^{bm}\, T^c_{md} J^d_R({x_2})\,=\nonumber \\
 &=&\int_{z,u} e^{ikz+ipu}\int_{{x_1}{x_2}}\left[
 f_i(z-{x_1})[S_z-S_{x_1}]^{ac}\,  f_j(u-{x_2})[S_u-S_{{x_2}}]^{bd}\, \hat\rho^d_{{x_2}}\hat\rho^c_{x_1}\right. \,+\nonumber \\
 &+& f_i(z-{x_1})[S_z-S_{x_1}]^{ac}\,  f_j(u-{x_2})[S_u\delta_{u{x_1}}-S_{{x_2}}\delta_{{x_2}{x_1}}]^{bm}\, T^c_{md} \hat\rho^d_{{x_2}}
\end{eqnarray}
Symmetrizing the operators within this amplitude, one recovers eq.(\ref{amplitude}). Our aim here is to obtain the fully ordered expression for the probability. We therefore square the amplitude first and then perform the full symmetrization between the operators $\hat \rho$ entering both the amplitude and its conjugate. 
Let us introduce some notations that  will help to make our expressions compact.
\beq
N_{z{x_1}\bar z {\bar x_1}}\,\equiv\,\vec f(z-{x_1})\cdot \vec f({\bar x_1}-\bar z)
\eeq
\beq
F^{ab}_{z{x_1}\bar z{\bar x_1}}\,\equiv\,N_{z{x_1}\bar z {\bar x_1}}\, [(S_z^\dagger-S_{x_1}^\dagger)(S_{\bar z}-S_{{\bar x_1}})]^{ab}\,,\ \ \ \ \ \ \ \ \ \ \ \
F^{ab}_{{x_1}{\bar x_1}} (k)\,\equiv\,\int_{z\bar z} e^{ik(z-\bar z)}\,F^{ab}_{z{x_1}\bar z{\bar x_1}}\,,\ \ \ \ \ \ \ \ \ \ \ 
F^{ab}_{{x_1}{\bar x_1}}(k)\,=\,F^{ba}_{{\bar x_1}{x_1}}(-k)
\eeq
\beq
G^{ab}_{z{x_1}\bar z{\bar x_1}}\,\equiv\,N_{z{x_1}\bar z {\bar x_1}}\, [S_{x_1}^\dagger\,(S_{\bar z}-S_{{\bar x_1}})]^{ab}\,,\ \ \ \ \ \ \ \ \ \ 
G^{ab}_{{x_1}{\bar x_1}} (k)\,\equiv\,\int_{z\bar z} e^{ik(z-\bar z)}\,G^{ab}_{z{x_1}\bar z{\bar x_1}}\,,\ \ \ \ \ \ \ \ \ \ 
M^{ab}_{z{x_1}\bar z{\bar x_1}}\,\equiv\,N_{z{x_1}\bar z {\bar x_1}}\, [S_z^\dagger\,(S_{\bar z}-S_{{\bar x_1}})]^{ab}
\eeq
Squaring the amplitude eq.(\ref{A}) we obtain
\begin{equation}
{dN\over d^2pd^2kd\eta d\xi}=\langle\Sigma^4+\Sigma^2+\Sigma^3\rangle _{P,T}
\end{equation}
with (sum over both the color and coordinate indices is implied)
\begin{eqnarray}
\Sigma^4&=& F_{{x_1}{\bar x_1}}^{ab}(k)\,F_{{{x_2}}{{\bar x_2}}}^{cd}(p) 
\, \hat\rho^c_{{x_2}}  \hat\rho^a_{{x_1}} \hat\rho^d_{{\bar x_2}} \hat\rho^b_{{\bar x_1}} \nonumber \\
\Sigma^3&=&\left[\int_{u{\bar u}} e^{ip(u-{\bar u})}\,F^{ab}_{u{\bar x_1}}(k)\, M_{u{x_2} {\bar u} {\bar x_2}}^{md}\,T^a_{cm}\,  
-\, F^{ab}_{{x_1}{\bar x_1}}(k)\,G^{md}_{{x_2}{\bar x_2}} (p)\, T^a_{cm}
\right]\,\hat\rho^c_{{x_2}}   \hat\rho^d_{{\bar x_2}} \hat\rho^b_{{\bar x_1}} \,+\, \nonumber \\
&+& 
\left[\int_{u{\bar u}} e^{ip(u-{\bar u})}\,F^{ab}_{{x_1}{\bar u}}(k)\, M_{{\bar u}{\bar x_2} u {x_2}}^{mc}\,T^b_{md}\,  
-\, F^{ab}_{{x_1}{\bar x_2}}(k)\,G^{mc}_{{\bar x_2}{x_2}} (-p)\, T^b_{md}
\right]\,\hat\rho^c_{{x_2}} \hat\rho^a_{{x_1}}   \hat\rho^d_{{\bar x_2}} \nonumber \\
\Sigma^2&=& \int_{u{\bar u}} e^{ip(u-{\bar u})}\,
F_{{x_1}{\bar x_1}}^{ab}(k)\,N_{u{x_2} {\bar u}{\bar x_2}}\,T^a_{cm}\,\left[(S^\dagger_u\delta_{u{x_1}}-S^\dagger_{{x_2}}\delta_{{x_1}{x_2}})(S_{\bar u}\delta_{{\bar u}{\bar x_1}}-S_{{\bar x_2}}\delta_{{\bar x_1}{\bar x_2}})
\right]^{mn}\,T^b_{nd}\,\hat\rho^c_{{x_2}}    \hat\rho^d_{{\bar x_2}} 
\end{eqnarray}
Following \cite{kl}, the operators are substituted by commuting $c$-number functions according to:
\beq
\hat \rho_{x_1}^a \rightarrow\, \rho_{x_1}^b\left[1\,-\, {g\over 2}\left(T^c{\delta\over\delta \rho^c_{x_1}}\right) \,+\,{g^2\over 12}\left(T^c{\delta\over\delta \rho^c_{x_1}}\right)^2\ldots
 \right]^{ba}
\eeq
After long algebra we obtain the fully ordered expression 
\begin{equation}
{dN\over d^2pd^2kd\eta d\xi}=\langle\sigma^4+\bar\sigma^3+\bar \sigma^2\rangle _{P,T}
\end{equation}
with
\beq
\sigma^4\,=\,[\rho F(k)\rho]\,[\rho F(p)\rho]
\eeq

\begin{eqnarray}
\bar \sigma^3&=&\sigma^3\,+\,g\Big({1\over 2} \rho_x F_{xy}(-k) \tilde\rho_y F_{yw}(-p)\rho_w
\,+\,{1\over 2} \rho_x F_{xy}(k) \tilde\rho_y F_{yw}(p)\rho_w\,+\,
{1\over 2} tr[\tilde\rho_x F_{xx}(k)]\,[\rho_y F_{yw}(p)\rho_w]\,+\nonumber \\
&&+{1\over 2} [\rho_x F_{xy}(k) \rho_y] \,tr[\tilde\rho_x F_{xx}(p)]\Big)\nonumber \\
\sigma^3&=&g\Big({1\over 2} \rho_x F_{xy}(k) \tilde\rho_y F_{yw}(-p)\rho_w\,+\,{1\over 2} \rho_x F_{xy}(-k) \tilde\rho_y F_{yw}(p)\rho_w\,+\,
 \rho_x F_{xy}(-k) \tilde\rho_y G_{yw}(p)\rho_w\,-\, \rho_x F_{xy}(k) \tilde\rho_y G_{yw}(-p)\rho_w \,-\nonumber \\
 &-&
 \int_{u{\bar u}} e^{ip(u-{\bar u})}\,[\rho_{\bar x_1}F_{{\bar x_1}u}(-k)\tilde\rho_{{x_2}}M_{u{x_2} {\bar u} {\bar x_2}}\rho_{{\bar x_2}} \,+\,\rho_{x_1} F_{{x_1}{\bar u}}(k)\tilde\rho_{{\bar x_2}} M_{{\bar u}{\bar x_2} u{x_2}}\rho_{{x_2}}]\Big)
\end{eqnarray}
\begin{eqnarray}
\bar \sigma^2&=&\sigma^2+g^2\Big(-{1\over 4} tr[\tilde\rho_x F_{xx}(k)T^a] [F_{xy}(-p)\rho_y]^a
-{1\over 12} tr[ F_{xx}(k)T^a] [\tilde\rho_x F_{xy}(-p)\rho_y]^a
-{1\over 12} tr[T^a F_{xx}(k) \tilde\rho_x]  [F_{xy}(-p)\rho_y]^a \nonumber \\
&+&{1\over 4}[\rho_x F_{xy}(k)]^a tr[\tilde\rho_y F_{yy}(p)T^a]
\,-\,{1\over 4}[\rho_x F_{xy}(-k)]^a tr[\tilde\rho_y F_{yy}(p)T^a]
\,+\,{1\over 4}tr[\tilde\rho_x F_{xy}(k) \tilde\rho_y F_{yx}(p)]\,+\nonumber \\
&+&{1\over 4}tr[F_{xx}(k)T^a] [\tilde\rho_x F_{xy}(p)\rho_y]^a
\,+\,{1\over 4}tr[\tilde\rho_x F_{xx}(k)] tr[\tilde\rho_y F_{yy}(p)]
\,+\,{1\over 12} [\rho_x F_{xy}(-k) \tilde\rho_y]^a tr[ T^aF_{yy}(p)]\,- \nonumber \\
&-&{1\over 12} [\rho_x F_{xy}(-k)]^a tr[ T^aF_{yy}(p) \tilde\rho_y]
\,+\, {1\over 12} tr[\tilde\rho_x F_{xx}(k) T^a][F_{xy}(p)\rho_y]^a
\,+\, {1\over 12} tr[T^a F_{xx}(k) \tilde\rho_x][F_{xy}(p)\rho_y]^a\,-\nonumber \\
&-&{1\over 12} [\rho_x F_{xy}(k)]^a tr[ T^aF_{yy}(p) \tilde\rho_y]
\,+\,{1\over 12} [\rho_x F_{xy}(k)\tilde\rho_y]^a tr[ T^aF_{yy}(p) ]
\,-\,{1\over 2} [\rho_x F_{xy}(-k)]^a tr[T^aG_{yy}(p)\tilde\rho_y]\,+\nonumber \\
&+&{1\over 2} tr[ T^a F_{xx}(k) \tilde\rho_x] [G_{xy}(p)\rho_y]^a
\,-\,{1\over 2} [\rho_x F_{xy}(k)]^a tr[T^aG_{yy}(-p)\tilde\rho_y]
\,+\,{1\over 2} tr[ T^a F_{xx}(k) \tilde\rho_x] [G_{xy}(-p)\rho_y]^a\,+\nonumber \\
&+&{1\over 2} \int_{u{\bar u}} e^{ip(u-{\bar u})}\,\Big(
[\rho_{\bar x_1} F_{{\bar x_1}u}(-k)]^a tr[T^aM_{u{x_2} {\bar u}{x_2}}\tilde\rho_{{x_2}}]\,-\,
 tr[T^aF_{u{x_2}}(k)\tilde\rho_{{x_2}}] [M_{u{x_2} {\bar u} {\bar x_2}}\rho_{{\bar x_2}}]^a\,-\nonumber \\
&-&tr[\tilde\rho_{x_1} F_{{x_1}{\bar u}}(k)T^a] [M_{{\bar u}{x_1}u{x_2}}\tilde\rho_{{x_2}}]^a\,+\,
[\rho_{x_1}F_{{x_1}{\bar u}}(k)]^a tr [T^aM_{{\bar u}{x_2}u{x_2}}\tilde\rho_{{x_2}}]
\Big)\Big)\nonumber \\ \nonumber \\
\sigma^2&=& g^2\Big[-{1\over 4} tr[\tilde\rho_x F_{xy}(k)\tilde\rho_y F_{yx}(-p)] 
\,-\,{1\over 2} tr[ F_{xy}(-k)\tilde\rho_y G_{yx}(p)\tilde\rho_x]\,+\,{1\over 2} tr[ F_{xy}(k)\tilde\rho_y G_{yx}(-p)\tilde\rho_x]\,+\nonumber \\
&+&{1\over 2}\int_{u{\bar u}} e^{ip(u-{\bar u})}\,\left( tr[\tilde\rho_{{\bar x_2}}F_{{\bar x_2}u}(-k)\tilde\rho_{{x_2}} M_{u{x_2} {\bar u}{\bar x_2}}] \,+\,
 tr[\tilde\rho_{{x_2}}F_{{x_2} {\bar u}}(k)\tilde\rho_{{\bar x_2}} M_{{\bar u}{\bar x_2} u{x_2}}] 
\right)\,+\nonumber \\
&+&{1\over 2}\int_{u{\bar u}} e^{ip(u-{\bar u})}\,\Big( N_{u{x_2}{\bar u}{\bar x_2}} tr[F_{{\bar u}u}(-k)\tilde\rho_{{x_2}} S^\dagger_u S_{\bar u} \tilde\rho_{{\bar x_2}}] \,-\,
N_{u{x_2}{\bar u}{\bar x_2}} tr[F_{{\bar x_2}u}(-k)\tilde\rho_{{x_2}} S^\dagger_u S_{{\bar x_2}} \tilde\rho_{{\bar x_2}}]  \,-\nonumber \\
&-& N_{u{x_2}{\bar u}{\bar x_2}} tr[F_{{\bar u}{x_2}}(-k)\tilde\rho_{{x_2}} S^\dagger_{{x_2}} S_{\bar u} \tilde\rho_{{\bar x_2}} ]\,+\,
N_{u{x_2}{\bar u}{\bar x_2}} tr[F_{\bar  {\bar x_1}{x_2}}(-k)\tilde\rho_{{x_2}} S^\dagger_{{x_2}} S_{{\bar x_2}} \tilde\rho_{{\bar x_2}}]  
\Big)\Big]
\end{eqnarray}
This can now be averaged over the projectile wave function with the help of classical ensemble averaging procedure
\begin{equation}
\langle ... \rangle_P=\int [d\rho] W_P[\rho] ...
\end{equation}

\section*{Acknowledgments}
AK thanks the Physics Department of the Ben Gurion University for hospitality when part of this work was done.
The work of AK is supported by DOE grant DE-FG02-92ER40716.
The work of ML is partially supported by the Marie Curie Grant  PIRG-GA-2009-256313.



\begin{thebibliography}{99}

\bibitem{cms} [CMS Collaboration], e-Pritn arXiv:1009.4122.

\bibitem{recent} 
S.M. Troshin, N.E. Tyurin, 
e-Print: arXiv:1009.5229; 
Igor M. Dremin, Victor T. Kim, 
e-Print: arXiv:1010.0918 [hep-ph];
I.O. Cherednikov, N.G. Stefanis, 
e-Print: arXiv:1010.4463;
K. Werner, Iu. Karpenko, T. Pierog, 
e-Print: arXiv:1011.0375 ;
I. Bautista, J.Dias de Deus, C. Pajares, 
e-Print: arXiv:1011.1870 
 

\bibitem{ddgjlv} A. Dumitru, K. Dusling, F. Gelis, J. Jalilian-Marian, T. Lappi and  R. Venugopalan, e-Print: arXiv:1009.5295 [hep-ph] 


\bibitem{rhicridge} B. Alver, et al., [PHOBOS Collaboration] Phys.Rev.Lett.104:062301,2010, e-Print: arXiv:0903.2811;
 B.I. Abelev, et al., [STAR Collaboration]Phys.Rev.C80:064912,2009, e-Print: arXiv:0909.0191.


\bibitem{rhicexplane}  N. Armesto, C. Salgado and U. Wiedemann, Phys.Rev.Lett. 83, 242301 (2004); S. A. Voloshin, Phys. Lett. B 632, 490 (2006); E. V. Shuryak, Phys. Rev. C 76, 047901 (2007);

\bibitem{glv} A. Dumitru, F. Gelis , L. McLerran  and R.Venugopalan,  Nucl.Phys.A810:91-108,2008.
e-Print: arXiv:0804.3858 [hep-ph]; K. Dusling,  F. Gelis, T. Lappi and R. Venugopalan,  Nucl.Phys.A836:159-182,2010.
e-Print: arXiv:0911.2720 [hep-ph] 

\bibitem{glasma} T. Lappi and L. McLerran
 Nucl.Phys.A772:200-212,2006, e-Print: hep-ph/0602189


\bibitem{classical} 
A. Kovner, L. D. McLerran and H. Weigert,
Published in Phys.Rev.D52:6231-6237,1995.
e-Print: hep-ph/9502289;
 Phys.Rev.D52:3809-3814,1995.
e-Print: hep-ph/9505320;
F. Gelis and R. Venugopalan, 
Nucl.Phys.A776:135-171,2006, e-Print: hep-ph/0601209; Nucl.Phys.A779:177-196,2006, e-Print: hep-ph/0605246

\bibitem{eikonal} T. Altinoluk, A. Kovner and J. Peressutti, 
 Nucl.Phys.A818:232-245,2009.
e-Print: arXiv:0810.4533 [hep-ph] 

\bibitem{mv} L. McLerran and R. Venugopalan,Phys.Rev.D49:2233-2241,1994.
e-Print: hep-ph/9309289; Phys.Rev.D49:3352-3355,1994.
e-Print: hep-ph/9311205

\bibitem{bubbles} A. Kovner and M. Lublinsky, 
 Phys.Rev.D71:085004,2005, e-Print: hep-ph/0501198;  A. Kovner, M. Lublinsky and U.A. Wiedemann,
 JHEP 0706:075,2007.
e-Print: arXiv:0705.1713 [hep-ph] 


\bibitem{jamal}
  J.~Jalilian-Marian and Y.~V.~Kovchegov,
  Phys.\ Rev.\ D {\bf 70}, 114017 (2004)
  [Erratum-ibid.\ D {\bf 71}, 079901 (2005)].

\bibitem{multigluons}
A. Kovner, M. Lublinsky,
JHEP 0611:083,2006.
e-Print: hep-ph/0609227

\bibitem{francois} F. Gelis, T. Lappi and R Venugopalan, Phys.Rev.D78:054020,2008.
e-Print: arXiv:0807.1306 [hep-ph] 


\bibitem{pomloops2} T. Altinoluk, A. Kovner and M. Lublinsky, 
JHEP 0903:110,2009.
e-Print: arXiv:0901.2560 [hep-ph];

%
%
\bibitem{baier} R. Baier, A. Kovner, M. Nardi and U. Wiedemann,
 Phys.Rev.D72:094013,2005.
e-Print: hep-ph/0506126

 
 
\bibitem{kl} A. Kovner and M. Lublinsky,
Phys.Rev.D72:074023,2005.
e-Print: hep-ph/0503155; 
Nucl.Phys.A767:171-188,2006.
e-Print: hep-ph/0510047

\bibitem{tuomas} T. Lappi , S. Srednyak and R. Venugopalan, 
JHEP 1001:066,2010; e-Print: arXiv:0911.2068 [hep-ph] 

\bibitem{bk} I. Balitsky, {\it Nucl. Phys.}  {\bf B463} 99 (1996);
 Y.~V.~Kovchegov,
  Phys.\ Rev.\ D {\bf 61}, 074018 (2000)

\bibitem{kevin} K. Dusling, talk at the EMMI workshop, GSI November 2010.

\bibitem{adrianjamal} A. Dumitru and J. Jalilian-Marian,
e-Print: arXiv:1001.4820 [hep-ph] 

\bibitem{dipole} A. Mueller,
Nucl. Phys. B335 115, 1990; {\it ibid}  B 415
373, 1994; {\it ibid}  B437, 107, 1995 [e-Print Archive:hep-ph/9408245];

\bibitem{jimwlk} J. Jalilian Marian, A. Kovner, A.Leonidov and H.
Weigert,
Nucl. Phys. B504, 415, 1997 [e-Print Archive: hep-ph/9701284]
Phys. Rev. D59, 014014, 1999 [e-Print Archive: hep-ph/9706377]
J. Jalilian Marian, A. Kovner and H. Weigert, Phys. Rev. D59, 014015, 1999  [e-Print Archive: hep-ph/9709432];
A. Kovner and J.G. Milhano, Phys. Rev. D61, 014012, 2000  [e-Print Archive: hep-ph/9904420].
 A. Kovner, J.G. Milhano and H. Weigert,
Phys. Rev. D62, 114005,2000  [e-Print Archive:hep-ph/0004014];  H. Weigert, Nucl. Phys. A703, 823, 2002[e-Print Archive:hep-ph/0004044];  E.~Iancu, A.~Leonidov and L.~D.~McLerran,
  Phys. Lett.  B510, 133, 2001 [e-Print Archive:hep-ph/0102009];  Nucl. Phys.A692, 583, 2001 [e-Print Archive: hep-ph/0011241]; E. Ferreiro, E. Iancu, A. Leonidov, L. McLerran, 
Nucl. Phys. A703, 489, 2002 [e-Print Archive:hep-ph/0109115].


\bibitem{yoshi} Y. Hatta, A. H. Mueller,
 Nucl.Phys.A789:285-297,2007, e-Print: arXiv:hep-ph/0702023
 
 \bibitem{kl1}  A. Kovner and M. Lublinsky
      JHEP 0503:001,2005, e-Print: arXiv:hep-ph/0502071
      
\bibitem{nestor} N. Armesto and J.G. Milhano, 
Phys.Rev.D73:114003,2006.
e-Print: hep-ph/0601132      
      
\bibitem{munier} A.H. Mueller and S. Munier,
 Phys.Rev.D81:105014,2010,
e-Print: arXiv:1002.4575 [hep-ph] 
 
%
%


\bibitem{LL}
  E.~Levin and M.~Lublinsky,
  Nucl.\ Phys.\  A {\bf 730}, 191 (2004)
  [arXiv:hep-ph/0308279];
  Phys.\ Lett.\  B {\bf 607}, 131 (2005)
  [arXiv:hep-ph/0411121].


\bibitem{klw}
  A.~Kovner, M.~Lublinsky and H.~Weigert,
  Phys.\ Rev.\  D {\bf 74}, 114023 (2006)
  [arXiv:hep-ph/0608258].

\bibitem{star} H. Agakishiev et.al. (STAR collaboration),
e-Print arXiv:1010.0690 

\end{thebibliography}
\end{document}